\documentclass[american]{article}
\usepackage[T1]{fontenc}
\usepackage[latin9]{inputenc}
\usepackage{geometry}
\usepackage{amsbsy}
\usepackage{amssymb}
\usepackage{graphicx}
\usepackage{esint}
\usepackage{babel}
\begin{document}

\title{Kinetic distance and kinetic maps from molecular dynamics simulation}

\author{Frank No\'{e}$^{1,*}$ and Cecilia Clementi$^{2,*}$}

\maketitle
$^{1}$: FU Berlin, Department of Mathematics, Computer Science and
Bioinformatics, Arnimallee 6, 14195 Berlin

$^{2}$: Rice University, Center for Theoretical Biological Physics,
and Department of Chemistry, Houston, TX 77005 

$*$: correspondence to frank.noe@fu-berlin.de or cecilia@rice.edu
\begin{abstract}
Characterizing macromolecular kinetics from molecular dynamics (MD)
simulations requires a distance metric that can distinguish slowly-interconverting
states. Here we build upon diffusion map theory and define a kinetic
distance for irreducible Markov processes that quantifies how slowly
molecular conformations interconvert. The kinetic distance can be
computed given a model that approximates the eigenvalues and eigenvectors
(reaction coordinates) of the MD Markov operator. Here we employ the
time-lagged independent component analysis (TICA). The TICA components
can be scaled to provide a kinetic map in which the Euclidean distance
corresponds to the kinetic distance. As a result, the question of
how many TICA dimensions should be kept in a dimensionality reduction
approach becomes obsolete, and one parameter less needs to be specified
in the kinetic model construction. We demonstrate the approach using
TICA and Markov state model (MSM) analyses for illustrative models,
protein conformation dynamics in bovine pancreatic trypsin inhibitor
and protein-inhibitor association in trypsin and benzamidine.
\end{abstract}

\section{Introduction}

Characterizing the metastable (long-lived) states, their equilibrium
probabilities and the transition rates between them are essential
aims of molecular dynamics (MD) simulation. Due to increasing availability
of high-performance computing resources for mass production of MD
data, there is an increasing interest in systematic and automatic
construction of models of the metastable dynamics, such as Markov
state models (MSMs) \cite{swope:jpcb:2004:markov-model-theory,noe:jcp:2007:markov-models,chodera:jcp:2007:automatic-state-decomposition,buchete-hummer:2008:coarse-master-equations,PrinzEtAl_JCP10_MSM1,BowmanPandeNoe_MSMBook}
and diffusion maps \cite{CoifmanLafon_PNAS05_DiffusionMaps,RohrdanzClementi_JCP134_DiffMaps}.
Alternative methods to analyze the space explored by MD simulations
include sketchmap \cite{Parrinello_sketchmap_PNAS2011}, PCA \cite{Altis:2008iv},
kernel-PCA \cite{Pozun:2012fo,antoniou2011toward}, and likelihood
maximization \cite{Peters:2007et}. See \cite{ClementiAnnualReview2012}
for a more extensive review. 

Critical component in any analysis of metastable dynamics from MD
data are (1) the choice of suitable coordinates and (2) a distance
metric on these coordinates, such that slowly-interconverting states
are distinguished. The first question has been answered: building
on conformation dynamics theory \cite{SchuetteFischerHuisingaDeuflhard_JCompPhys151_146,SarichNoeSchuette_MMS09_MSMerror,PrinzEtAl_JCP10_MSM1,NoeNueske_MMS13_VariationalApproach}
it has been shown that the eigenfunctions of the backward Markov propagator
underlying the MD are the optimal reaction coordinates: Projecting
the dynamics upon these eigenfunctions will give rise to a maximum
estimate of the timescales \cite{PrinzChoderaNoe_PRX14_RateTheory,NoeNueske_MMS13_VariationalApproach,NueskeEtAl_JCTC14_Variational}
and an optimal separation of metastable states. A number of approaches
are available to approximate these reaction coordinates from MD data:
Diffusion maps \cite{RohrdanzClementi_JCP134_DiffMaps}, Markov state
models \cite{PrinzEtAl_JCP10_MSM1} and Markov transition models \cite{WuNoe_JCP15_GMTM},
TICA \cite{PerezEtAl_JCP13_TICA,SchwantesPande_JCTC13_TICA} and kernel
TICA \cite{SchwantesPande_JCTC15_kTICA}. The variational approach
for conformation dynamics (VAC) \cite{NoeNueske_MMS13_VariationalApproach,NueskeEtAl_JCTC14_Variational}
is a generalization to all aforementioned models except for diffusion
maps and describes a general approach to combine and parametrize basis
functions so as to optimally define the true eigenfunctions of the
backward propagator.

The second question, i.e. the choice of a distance metric has been
answered in the context of diffusion maps. Diffusion maps model the
observed data as emerging from a diffusion process \cite{CoifmanLafon_PNAS05_DiffusionMaps,NadlerLafonCoifmanKevrikidis_NIPS05_DiffMaps}.
This approach has been used to model molecular conformation dynamics
\cite{RohrdanzClementi_JCP134_DiffMaps} and to guide further sampling
\cite{PretoClementi_PCCP14_AdaptiveSampling}. The diffusion distance,
introduced in \cite{NadlerLafonCoifmanKevrikidis_NIPS05_DiffMaps}
defines a distance metric that measures how slowly conformations interconvert.
However, the definition of the diffusion distance is not limited to
diffusion processes or even to reversible Markov processes. 

Here we define the kinetic distance for Markov processes that have
a unique equilibrium distribution, such as MD simulation. For reversible
Markov processes the kinetic distance takes the same simple form as
the diffusion distance and can be computed from any model that provides
an approximation to the eigenvalues and eigenfunctions of the MD backward
propagator. Here we employ the time-lagged independent component analysis
(TICA) \cite{PerezEtAl_JCP13_TICA,SchwantesPande_JCTC13_TICA} which
approximates these eigenvalues and eigenfunctions in terms of a linear
combinations of molecular coordinates such as atomic positions, distances
or angles. While TICA does not provide the best possible approximation
to the true eigenfunctions, it can be directly applied to molecular
dynamics data, requires few modeling decisions to be made, and is
readily available in the Markov modeling packages EMMA \cite{SenneSchuetteNoe_JCTC12_EMMA1.2}
(www.pyemma.org) and MSMbuilder \cite{BeauchampEtAl_MSMbuilder2}.

Using the TICA eigenvalues and eigenvectors we employ the kinetic
distance as a distance metric. As in diffusion maps, the approximate
eigenvectors can be rescaled so as to define a kinetic map in which
the Euclidean distance is equivalent to the kinetic distance. This
provides an optimal space to perform clustering, Markov modeling,
and diffusion map, or to use the Variational Approach for conformation
dynamics and performing other analyses of the metastable dynamics.
The kinetic map gives a clear answer to the previously arbitrary decision
on how many dimensions should be kept in the TICA transformation:
In principle all dimensions are kept, but as a result of the scaling
the dimensions with small eigenvalues contribute very little to the
kinetic distance. In order to reduce the computational cost, we can
choose to keep only the eigenvectors that account for a certain fraction
of the total variation in kinetic distance, as it is commonly practiced
in principal component analysis \cite{Hotelling_JEduPsy24_441,Amadei_Proteins17_412}.

We demonstrate the approach using TICA and MSM analyses for illustrative
models, protein conformation dynamics in bovine pancreatic trypsin
inhibitor and protein-inhibitor association in trypsin and benzamidine.
By invoking the Variational Principle \cite{NoeNueske_MMS13_VariationalApproach}
as a means of model comparison, it is shown that the kinetic map generally
provides better MSMs than by using unscaled TICA with manually selected
dimension.

\section{Theory}

\subsection{Diffusion distance and kinetic distance}

Suppose we have a dynamical system (here molecular dynamics) with
a state space $\Omega$. $\Omega$ formally includes positions, momenta
and if needed other state variables such as the simulation box size,
although we will later make approximations such as working with only
a subset of the configuration coordinates. Our dynamics generates
a time sequence of states $\mathbf{x}\in\Omega$ by means of a Markovian
algorithm, i.e. some implementation of time-step integrator, thermostat,
etc. that allows us to compute the system state in the subsequent
time step as a function of the current state. In this framework there
is a transition density $p_{\tau}(\mathbf{y}\mid\mathbf{x})$, the
probability density of finding the system at state $\mathbf{y}$ at
time $\tau$ given that we have started it at state $\mathbf{x}$
at time 0. Given these preliminaries we can write the propagation
of a probability density of states $\rho_{t}(\mathbf{x})$ in time
as:
\begin{eqnarray}
\rho_{t+\tau}(\mathbf{y}) & = & \int_{\mathbf{x}\in\Omega}\rho_{t}(\mathbf{x})\,p_{\tau}(\mathbf{y}\mid\mathbf{x})\,\mathrm{d}\mathbf{x}\label{eq:propagator_1}\\
 & = & \mathcal{P}\circ\rho_{t}(\mathbf{x})
\end{eqnarray}
where $\mathcal{P}$ is the dynamical propagator, i.e. the Markov
operator describing the action of the integral in (\ref{eq:propagator_1}).
Finally, we require that there is a unique equilibrium distribution
(usually the Boltzmann distribution) defined by:
\begin{equation}
\pi(\mathbf{x})=\mathcal{P}\circ\pi(\mathbf{x})
\end{equation}
The above requirements are minimal as they are fulfilled by almost
every implementation of molecular dynamics. There is an alternative
description to (\ref{eq:propagator_1}) by using the backward propagator,
$\mathcal{T}$, (also known as transfer operator \cite{SchuetteFischerHuisingaDeuflhard_JCompPhys151_146}),
which propagates the weighted densities $v_{t}(\mathbf{x})=\rho_{t}(\mathbf{x})/\pi(\mathbf{x})$.
Thus:
\begin{eqnarray}
v_{t+\tau}(\mathbf{y}) & = & \frac{1}{\pi(\mathbf{y})}\int_{\mathbf{y}}p_{\tau}(\mathbf{y}\mid\mathbf{x})\,\pi(\mathbf{x})\,v_{t}(\mathbf{x})\,\mathrm{d}\mathbf{x}\\
 & = & \mathcal{T}\circ v_{t}(\mathbf{x}).
\end{eqnarray}

We define the kinetic distance $D_{\tau}(\mathbf{x}_{1},\,\mathbf{x}_{2})$
as the diffusion distance introduced in \cite{NadlerLafonCoifmanKevrikidis_NIPS05_DiffMaps}
in the more general context of the above irreducible Markov processes.
$D_{\tau}(\mathbf{x}_{1},\,\mathbf{x}_{2})$ is a distance measure
between any of the two states $\mathbf{x}_{1},\mathbf{x}_{2}\in\Omega$
and parametrically depends on a lag time $\tau$. $D_{\tau}(\mathbf{x}_{1},\,\mathbf{x}_{2})$
is defined as the distance between the system's probability densities
at time $\tau$, given that we have initialized the system either
at state $\mathbf{x}_{1}$ or at state $\mathbf{x}_{2}$ at time $0$:
\begin{eqnarray}
D_{\tau}^{2}(\mathbf{x}_{1},\,\mathbf{x}_{2}) & = & \left\Vert p_{\tau}(\mathbf{y}\mid\mathbf{x}_{1})-p_{\tau}(\mathbf{y}\mid\mathbf{x}_{2})\right\Vert _{\pi^{-1}}^{2}\label{eqn:diffdist1}\\
 & = & \int_{\mathbf{y}\in\Omega}\frac{|p_{\tau}(\mathbf{y}\mid\mathbf{x}_{1})-p_{\tau}(\mathbf{y}\mid\mathbf{x}_{2})|^{2}}{\pi(\mathbf{y})}\:\mathrm{d}\mathbf{y}.
\end{eqnarray}
The above definition is identical to the concept of the diffusion
distance except for the fact that in principle we allow $\mathbf{x}_{1}$,
$\mathbf{x}_{2}$ and $\mathbf{y}$ to be not only positions but also
momenta and other state variables. This modification is needed in
order to derive expressions of the kinetic distance that also apply
to nonreversible dynamics. Furthermore, the original name diffusion
distance comes from the fact that it has been derived in a context
where the dynamics $p_{\tau}(\mathbf{y}\mid\mathbf{x})$ come from
a diffusion process. Here we will apply (\ref{eqn:diffdist1}) to
a wider class of Markovian dynamics, and therefore use the term kinetic
distance rather than diffusion distance throughout the article.

\subsection{Spectral decomposition}

In order to derive practically useful expressions of the kinetic distance,
we need to conduct a spectral decomposition of (\ref{eq:propagator_1}).
Given the propagator $\mathcal{P}$ with eigenfunctions $\phi_{i}$
and the backward propagator $\mathcal{T}$ with eigenfunctions $\psi_{i}$,
and we suppose that our propagator has a number of $n$ discrete eigenpairs.
The rest of the spectrum is bounded by the ball with radius $|\lambda_{n+1}|$
and it will be called $\mathcal{P}^{\mathrm{fast}}$. We can write
the propagation of densities $\rho$ as follows:
\begin{equation}
\rho_{t+\tau}(\mathbf{y})=\sum_{j=1}^{n}\lambda(\tau)\langle\psi_{j}(\mathbf{x})\mid\rho_{t}(\mathbf{x})\rangle\phi_{j}(\mathbf{y})+\mathcal{P}^{\mathrm{fast}}(\tau)\circ\rho_{t}(\mathbf{x}).
\end{equation}
The norm of all eigenvalues decay exponentially with increasing lag
time. We suppose that we operate at a lag time $\tau$ such that $|\lambda_{n+1}|\approx0$,
and thus $\mathcal{P}^{\mathrm{fast}}(\tau)\circ\rho_{t}(\mathbf{x})$
is approximately 0 everywhere. Then we can effectively describe the
dynamics as:
\begin{equation}
\rho_{t+\tau}(\mathbf{y})=\sum_{j=1}^{n}\lambda(\tau)\langle\psi_{j}(\mathbf{x})\mid\rho_{t}(\mathbf{x})\rangle\phi_{j}(\mathbf{y})
\end{equation}
and for the choice $\rho_{t}(\mathbf{x})=\delta(\mathbf{x})$ this
becomes:
\begin{equation}
\rho_{t+\tau}(\mathbf{y})=\sum_{j=1}^{n}\lambda(\tau)\psi_{j}(\mathbf{x})\phi_{j}(\mathbf{y}).\label{eqn:spectral_decomposition}
\end{equation}

We sort the eigenvalues by non-increasing norm:
\begin{equation}
\lambda_{1}=1>|\lambda_{2}|\ge|\lambda_{3}|\ge\,...\,\ge|\lambda_{n}|\label{eq:eval_sort}
\end{equation}
The eigenfunction $\varphi_{1}(\mathbf{x})=\pi(\mathbf{x})$ corresponding
to the eigenvalue $\lambda_{1}=1$ is given by the Boltzmann equilibrium
distribution. 

Note that we have not made any restriction with respect to the reversibility
of the dynamics as it is usually done in Markov modeling \cite{PrinzEtAl_JCP10_MSM1}.
Thus the eigenvalues $\lambda(\tau)$ can either be all real-valued,
or consist of a mix of real eigenvalues and complex conjugate pairs.
For the calculation of the kinetic distance these cases need to be
treated slightly differently.

\subsection{Computing kinetic distances and kinetic maps}

\paragraph{Nonreversible dynamics}

Inserting (\ref{eqn:spectral_decomposition}) into (\ref{eqn:diffdist1})
yields:
\begin{eqnarray}
D_{\tau}^{2}(\mathbf{x}_{1},\,\mathbf{x}_{2}) & = & \left\Vert \sum_{j=1}^{n}\lambda_{j}(\tau)\left(\psi_{j}(\mathbf{x}_{1})-\psi_{j}(\mathbf{x}_{2})\right)\phi_{j}(\mathbf{y})\right\Vert _{\pi^{-1}}^{2}
\end{eqnarray}
As a result of $\psi_{1}(\mathbf{x})\equiv1$, the first term with
$j=1$ disappears. Evaluating the square norm leads to the following
double sum:
\begin{equation}
D_{\tau}^{2}(\mathbf{x}_{1},\,\mathbf{x}_{2})=\sum_{j=2}^{n}\sum_{k=2}^{n}\left(\psi_{j}(\mathbf{x}_{1})-\psi_{j}(\mathbf{x}_{2})\right)\lambda_{j}\langle\phi_{j}\mid\phi_{k}\rangle_{\pi^{-1}}\lambda_{k}\left(\psi_{k}(\mathbf{x}_{1})-\psi_{k}(\mathbf{x}_{2})\right)\label{eq:diffdist_nonrev}
\end{equation}
The above expression is applicable to any Markov process with a unique
stationary distribution. Unfortunately (\ref{eq:diffdist_nonrev})
requires both the propagator and the backward propagator eigenvectors,
and there are only few models for nonreversible dynamics that provide
both sets of eigenvectors. Markov state models provide both eigenvectors
(as left and right eigenvectors of the transition matrix). Moreover
it is possible to design Markov transition models \cite{WuNoe_JCP15_GMTM}
such that both sets of eigenvectors can be computed. 

For nonreversible dynamics, there can be complex eigenvalues and eigenvectors
which come in complex conjugate pairs $(\lambda_{j},\phi_{j},\psi_{j})$
and $(\bar{\lambda}_{j},\bar{\phi}_{j},\bar{\psi}_{j})$. These are
handled as follows: A complex conjugate pair is rewritten into a real
pair of eigenvalues/eigenvectors by separating their real and imaginary
parts:
\begin{eqnarray}
\lambda_{j},\bar{\lambda}_{j} & \rightarrow & \mathrm{Re}(\lambda_{j}),\:\mathrm{Im}(\lambda_{j})\\
\phi_{j},\bar{\phi}_{j} & \rightarrow & \mathrm{Re}(\phi_{j}),\:\mathrm{Im}(\phi_{j})\\
\psi_{j},\bar{\psi}_{j} & \rightarrow & \mathrm{Re}(\psi_{j}),\:\mathrm{Im}(\psi_{j})
\end{eqnarray}
Note that the sign is irrelevant, so it doesn't matter if we use $(\lambda_{j},\phi_{j},\psi_{j})$
or $(\bar{\lambda}_{j},\bar{\phi}_{j},\bar{\psi}_{j})$ for the imaginary
part. The transformed eigenvalue/vector set which is still the same
rank but now real-valued is inserted into Eq. (\ref{eq:diffdist_nonrev})
to compute the diffusion distance.

\paragraph{Reversible dynamics}

For reversible dynamics the detailed balance equations hold. In that
case every pair of points $\mathbf{x}$, $\mathbf{y}$ satisfies:
\begin{equation}
\pi(\mathbf{x})p_{\tau}(\mathbf{y}\mid\mathbf{x})=\pi(\mathbf{y})p_{\tau}(\mathbf{x}\mid\mathbf{y}).\label{eq:DB}
\end{equation}
In this case all eigenvalues $\lambda_{i}(\tau)$ are real and can
be associated with relaxation rates $\kappa_{i}$ and timescales $t_{i}$
as:
\begin{equation}
\lambda_{i}(\tau)=\mathrm{e}^{-\tau\kappa_{i}}=\mathrm{e}^{-\tau/t_{i}}.
\end{equation}
Moreover, all associated eigenvectors $\psi_{i},\phi_{i}$ are real
and related by:
\begin{equation}
\phi_{i}(\mathbf{x})=\pi(\mathbf{x})\psi_{i}(\mathbf{x}).\label{eq:eigenvectors-l-r-rev}
\end{equation}

Given Eq. (\ref{eq:eigenvectors-l-r-rev}), we have $\langle\phi_{j}\mid\phi_{k}\rangle_{\pi^{-1}}=\delta_{jk}$
where $\delta_{jk}$ is the Dirac delta. Thus, only the $j=k$ terms
survive in Eq. (\ref{eqn:spectral_decomposition}), which simplifies
to:
\begin{equation}
D_{\tau}^{2}(\mathbf{x}_{1},\,\mathbf{x}_{2})=\sum_{j=2}^{n}\left(\lambda_{j}\psi_{j}(\mathbf{x}_{1})-\lambda_{j}\psi_{j}(\mathbf{x}_{2})\right)^{2}.\label{eq:diffdist_rev}
\end{equation}
Which is identical to the expression for the diffusion distance \cite{NadlerLafonCoifmanKevrikidis_NIPS05_DiffMaps}.
Note, however, we can apply it to any dynamics which is reversible
in the state space used in our model. Diffusion maps employ Smoluchowski
or Brownian dynamics which are reversible. Langevin dynamics, which
is more commonly used for thermostatting MD simulations fulfills a
generalized detailed balance in phase space with respect to momentum
inversion. Moreover, it has been recently shown that Langevin dynamics
fulfills (\ref{eq:DB}) in position space when integrating over momenta
\cite{BittracherKoltaiJunge_arxiv14_SpatialTransferOperators}. Since
kinetic models for estimating eigenvalues and eigenvectors are usually
estimated in position space (or a subset thereof), Langevin dynamics
is consistent with the reversible diffusion distance Eq. (\ref{eq:diffdist_rev}).

Based on Eq. (\ref{eq:diffdist_rev}), we can define the weighted
coordinates
\begin{equation}
\tilde{\psi}_{i}(\mathbf{x})=\lambda_{i}(\tau)\psi_{i}(\mathbf{x}),
\end{equation}
which define the kinetic map:
\begin{equation}
\tilde{\Psi}=\left(\tilde{\psi}_{1},\,...,\,\tilde{\psi}_{n}\right)^{\top}.
\end{equation}
The kinetic map is a new set of coordinates in which data points $\mathbf{x}$
have been transformed such that their Euclidean distance corresponds
to their diffusion distance:
\begin{equation}
D_{\tau}^{2}(\mathbf{x}_{1},\,\mathbf{x}_{2})=\left\Vert \widetilde{\Psi}(\mathbf{x}_{1})-\widetilde{\Psi}(\mathbf{x}_{2})\right\Vert ^{2}\label{eq:kinetic_map}
\end{equation}
When the transition density $p(\mathbf{y}|\mathbf{x})$ originates
from a diffusion process, then $\widetilde{\Psi}(\mathbf{x})$ is
called a diffusion map \cite{NadlerLafonCoifmanKevrikidis_NIPS05_DiffMaps}. 

If the number of available eigenvectors $n$ is large, then evaluating
distances in (\ref{eq:kinetic_map}) is computationally costly. However,
this is often unnecessary because many eigenvalues may be small and
the corresponding dimensions in the kinetic map contribute little
to the overall distance. We can employ an approach commonly used in
principal component analysis with geometric distances: We compute
the variance of the diffusion distance along each coordinate:
\begin{equation}
\langle\tilde{\psi}_{i},\tilde{\psi}_{i}\rangle_{\pi}=\lambda_{i}^{2}(\tau)=\mathrm{e}^{-2\tau/\kappa_{i}}
\end{equation}
and the cumulative variance fraction (remember the eigenvalues are
sorted by decreasing norm) by:
\begin{equation}
c_{k}=\frac{\sum_{i=2}^{k}\lambda_{i}^{2}(\tau)}{\sum_{i=2}^{n}\lambda_{i}^{2}(\tau)}
\end{equation}
We can then decide to truncate the distance after $k=K$ terms where
a certain cumulative variance threshold (e.g. 95\%). The approximate
diffusion distance is then given by:
\begin{equation}
D_{\tau}(\mathbf{x}_{1},\,\mathbf{x}_{2})\approx\sqrt{\sum_{i=2}^{K}\left(\tilde{\psi}_{i}(\mathbf{x}_{1})-\tilde{\psi}_{i}(\mathbf{x}_{2})\right)^{2}}.
\end{equation}

\subsection{Algorithmic approach using TICA}

In order to apply the kinetic distance and kinetic maps, we need an
algorithm that will approximate the eigenvalues $\lambda_{i}$ and
eigenfunctions $\psi$. As described in the introduction a number
of methods are available for this. Here we choose to use the TICA
method as implemented in pyEMMA. Although TICA only provides only
a rather rough approximation to eigenvalues and eigenfunctions, it
is very easy to use, very robust, and has few parameters to choose.
Given MD data, we chose a (usually large) set of input coordinates
$\{r_{i}(t)\}$, such as Cartesian coordinates (if there is a reference
to orient the solute molecule(s) to) or internal coordinates (inter-residue
distances, rotamer dihedral angles, etc.). We define the mean-free
coordinates:
\begin{equation}
y_{i}(t)=r_{i}(t)-\langle r_{i}(t)\rangle_{t}\label{eq:tica_basis_set}
\end{equation}
and compute the covariance matrix and time-lagged covariance matrix
for a given lag time $\tau$:
\begin{eqnarray}
c_{ij}(0) & = & \langle y_{i}(t)y_{j}(t)\rangle_{t}\\
c_{ij}(\tau) & = & \langle y_{i}(t)y_{j}(t+\tau)\rangle_{t}
\end{eqnarray}
In practice, means and covariances are computed by their empirical
estimators, and the time-lagged covariance matrix is symmetrized.
We then solve the generalized eigenvalue problem:
\begin{equation}
\mathbf{C}(\tau)\mathbf{r}_{i}=\mathbf{C}(0)\mathbf{r}_{i}\hat{\lambda}_{i}(\tau).\label{eq:gev}
\end{equation}
We have the following approximations for the eigenvalues and eigenfunctions
of the backward propagator:
\begin{eqnarray}
\hat{\lambda}_{i} & \lesssim & \lambda_{i}\label{eq:tica_eig_estimate}\\
\hat{\psi}_{i}=\sum_{j}r_{i,j}y_{j} & \approx & \psi_{i}\label{eq:tica_eigv_estimate}
\end{eqnarray}
The approach above is a special case of the Variational Approach of
conformation dynamics \cite{NoeNueske_MMS13_VariationalApproach,NueskeEtAl_JCTC14_Variational},
using the choice (\ref{eq:tica_basis_set}) as a basis set. According
to the variational principle \cite{NoeNueske_MMS13_VariationalApproach},
the eigenvalues $\hat{\lambda}_{i}$ will only be exactly $\lambda_{i}$
if $\hat{\psi}_{i}=\psi_{i}$ and otherwise will be underestimated.
For finite data there are statistical errors on top of these approximation
errors, and therefore the underestimation is denoted as an approximation
in Eq. (\ref{eq:tica_eig_estimate}).

We then apply the following approach:
\begin{enumerate}
\item Perform TICA and compute the eigenvectors $\hat{\psi}_{i}$ and eigenvalues
$\hat{\lambda}_{i}$. 
\item Define the \textbf{kinetic map} by scaling all coordinates as
\begin{equation}
\tilde{\psi}_{i}=\lambda_{i}(\tau)\psi_{i}
\end{equation}

\item The kinetic distance is defined by
\begin{equation}
D_{\tau}(\mathbf{x}_{1},\mathbf{x}_{2})=\left\Vert \tilde{\psi}_{i}(\mathbf{x})-\tilde{\psi}_{i}(\mathbf{x})\right\Vert _{2}
\end{equation}

\end{enumerate}
Note that TICA already projects the stationary eigenvector $\psi_{1}$
out as a consequence of its construction with mean-free coordinates
(\ref{eq:tica_basis_set}). As a result, all approximated eigenvectors
and eigenvalues are kept and the diffusion distance (\ref{eq:diffdist_rev})
runs through all terms $1,\,...,\,n$.

\section{Applications}

We demonstrate the behavior of the proposed kinetic distance using
time-series of two pedagogical and two real molecular systems. All
analyses are run with pyEMMA (www.pyemma.org).

In all examples, we conduct the following data analysis:
\begin{enumerate}
\item Transformation of the input coordinates to an approximation of eigenvalues
and eigenfunctions $(\lambda_{2},\,\psi_{2}),\,...,\,(\lambda_{n},\,\psi_{n})$
using TICA.
\item Cluster discretization of (i) the transformed space $\boldsymbol{\Psi}$,
(ii) dimension-reduced versions of it (TICA projections) as previously
practiced \cite{PerezEtAl_JCP13_TICA,SchwantesPande_JCTC13_TICA},
and (iii) the full-dimensional kinetic map (scaled TICA transformation)
proposed here. We use the same clustering method and equal number
of clusters for the same system to get comparable results.
\item Compute Markov models using these different discretization and compare
them using the variational principle \cite{NoeNueske_MMS13_VariationalApproach}.
\end{enumerate}
In order to compare different Markov models we employ the variational
principle of conformation dynamics \cite{NoeNueske_MMS13_VariationalApproach},
which states that the approximated eigenvalues computed via (\ref{eq:gev})
underestimate the true eigenvalues: 
\begin{equation}
\hat{\lambda}_{i}(\tau)\le\lambda_{i}(\tau)
\end{equation}
and that equality is only obtained when the corresponding eigenfunction
is correct ($\hat{\psi}_{i}\equiv\psi_{i}$). As a consequence, we
can make the same statement about approximated relaxation timescales:
\begin{equation}
\hat{t}_{i}(\tau)\le t_{i}(\tau).
\end{equation}
Straightforward fitness functions to compare Markov models obtained
in different ways are then the partial eigensum (sum over the first
$m$ eigenvalues) or the partial timescales sum. Here we use the mean
relaxation timescale:
\begin{equation}
MRT(m)=\frac{1}{m}\sum_{i=2}^{m}\hat{t}_{i}(\tau)\le\frac{1}{m}\sum_{i=2}^{m}t_{i}(\tau)\label{eq:timescale_sum}
\end{equation}
As we are doing only one TICA projection per molecular system we do
not use the TICA timescales but rather the final MSM timescales in
(\ref{eq:timescale_sum}). This will compare the quality of the different
projections / metrics (full TICA, truncated TICA or scaled TICA /
kinetic map) with respect to their ability to resolve the metastable
dynamics using a fixed clustering approach.

\textbf{Two-state Hidden Markov Models}

We start by illustrating two pedagogical examples that are realized
by Hidden Markov models (HMMs) with Gaussian output. In the first
example, our approach works especially well and in the second example
it fails as the result of a poor TICA approximation. 

\begin{figure}
\noindent \begin{centering}
a)\includegraphics[width=0.5\columnwidth]{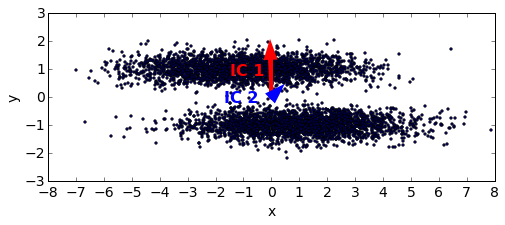}
\par\end{centering}

\noindent \begin{centering}
b)\includegraphics[width=0.5\columnwidth]{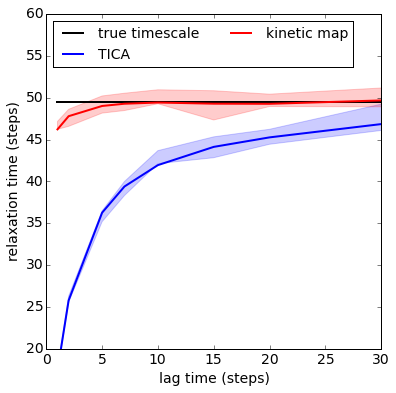}
\par\end{centering}

\protect\caption{\label{fig:hmm1}Comparison of Markov models using scaled and unscaled
TICA. (a) The data was generated by a two-state Hidden Markov model
with Gaussian output distributions. (b) Implied relaxation timescales
using scaled and unscaled TICA followed by regular-space clustering
with 28 clusters each. }
\end{figure}

In order to have a known reference for the timescales we fix a transition
matrix between two metastable states: 
\[
P=\left[\begin{array}{cc}
0.99 & 0.01\\
0.01 & 0.99
\end{array}\right]
\]
which leads to a single relaxation timescale of $t_{2}=49.5$ steps.
In order to generate coordinates we let each of the two states sample
from two-dimensional Gaussian distributions using means $\mu_{i}$
and covariance matrices $\Sigma_{i}$ given by:

\begin{eqnarray*}
\mu_{1}=\left(\begin{array}{c}
-1\\
1
\end{array}\right) &  & \Sigma_{1}=\left(\begin{array}{cc}
0.3^{2}\\
 & 2^{2}
\end{array}\right)\\
\mu_{2}=\left(\begin{array}{c}
1\\
-1
\end{array}\right) &  & \Sigma_{2}=\left(\begin{array}{cc}
0.3^{2}\\
 & 2^{2}
\end{array}\right)
\end{eqnarray*}
In order to analyze systematic rather than statistical behavior we
simulated this model for an extensive 250,000 steps starting from
state 1. In every simulation step we take a step using the transition
matrix $P$ (on average transitioning to the other state every 100
steps), and generate a point from Gaussian distribution 1 or 2, depending
which state we are in. Fig. \ref{fig:hmm1}a shows a scatterplot for
the resulting simulation. Additional details on how to construct a
HMM can be found in \cite{Rabiner_IEEE89_HMM}.

In this example, TICA works especially well. Fig. \ref{fig:hmm1}a
shows that while the main geometric variance is along the $y$-direction
(PCA would find that as principal component), the slow process is
along the $x$-direction. Indeed the first independent component (IC)
points exactly along the $x$-direction, while the second IC points
towards a mixtures of $x$ and $y$ directions (note that the ICs
are not orthogonal in Cartesian space but rather in the space weighted
by the equilibrium distribution). The arrows are drawn proportional
to the TICA eigenvalues which are $\lambda_{1}\approx0.75$ and $\lambda_{2}\approx0$. 

We compare two metrics: Euclidean distance in the two-dimensional
TICA space $\boldsymbol{\Psi}=(\hat{\psi}_{1},\hat{\psi}_{2})^{\top}$
and the kinetic map $\widetilde{\boldsymbol{\Psi}}=(\hat{\lambda}_{2}\hat{\psi}_{1},\hat{\lambda}_{2}\hat{\psi}_{2})^{\top}$.
We conduct regular space clustering \cite{PrinzEtAl_JCP10_MSM1} using
27 clusters in both cases. Fig. \ref{fig:hmm1}b shows the mean relaxation
timescales as a function of the lag time for the two cases. It is
seen that Euclidean distance in the kinetic map performs much better
than the unscaled TICA space. The reason for this behavior is that
the second eigenvalue is about zero, and thus the kinetic map effectively
ignores $\hat{\psi}_{2}$. Hence, in the kinetic map version, the
27 clusters perform a fine discretization of the slow reaction coordinate,
which has been accurately found by TICA. In contrast, in the unscaled
TICA version, the 27 clusters are scattered over a two-dimensional
space, leading to a much poorer resolution of the reaction coordinate
$x$. Thus, the kinetic map is better because TICA has already identified
the right coordinates - all we needed to do was to adjust the metric.

Let us look at a counterexample, where the current approach fails.
Although the theory of kinetic distance and kinetic map is correct,
the crucial point is that we have to be able to generate a sufficiently
good approximation of eigenfunctions and eigenvalues $(\lambda_{i},\psi_{i})$
in order to apply it. TICA does that by finding a linear combination
of input coordinates and will fail when the true eigenfunctions $\psi_{i}$
are still highly nonlinear in these coordinates. So let's design a
pathological example. We use the four-state transition matrix:
\[
P=\left[\begin{array}{cccc}
0.9 & 0.1\\
0.1 & 0.89 & 0.01\\
 & 0.01 & 0.89 & 0.1\\
 &  & 0.1 & 0.9
\end{array}\right]
\]
and define Gaussian distributions that output into two-dimensional
Cartesian space:
\begin{eqnarray*}
\mu_{1}=\left(\begin{array}{c}
-4\\
0
\end{array}\right) &  & \Sigma_{1}=\left(\begin{array}{cc}
0.2^{2}\\
 & 1
\end{array}\right)\\
\mu_{2}=\left(\begin{array}{c}
-1\\
0.5
\end{array}\right) &  & \Sigma_{2}=\left(\begin{array}{cc}
1\\
 & 0.1^{2}
\end{array}\right)\\
\mu_{3}=\left(\begin{array}{c}
1\\
0.5
\end{array}\right) &  & \Sigma_{3}=\left(\begin{array}{cc}
1\\
 & 0.1^{2}
\end{array}\right)\\
\mu_{4}=\left(\begin{array}{c}
4\\
0
\end{array}\right) &  & \Sigma_{4}=\left(\begin{array}{cc}
0.2^{2}\\
 & 1
\end{array}\right)
\end{eqnarray*}

The simulation is run again for 250,000 steps and the coordinates
from the TICA transformation or the kinetic map are discretized with
25 clusters using $k$-means.

\begin{figure}
\noindent \begin{centering}
a)\includegraphics[width=0.5\columnwidth]{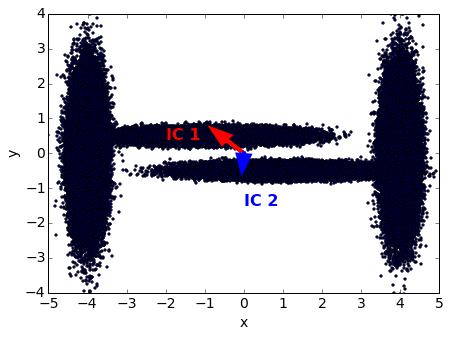} 
\par\end{centering}

\noindent \begin{centering}
b)\includegraphics[width=0.5\columnwidth]{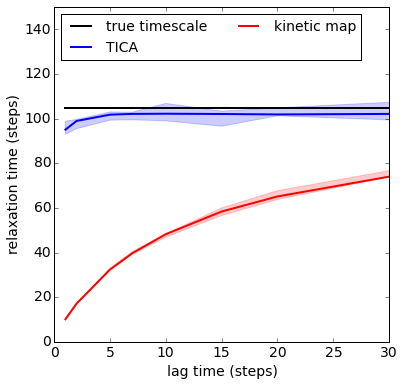}
\par\end{centering}

\protect\caption{\label{fig:hmm2}Comparison of Markov models using scaled and unscaled
TICA. (a) The data was generated by a two-state Hidden Markov model
with Gaussian output distributions. (b) Implied relaxation timescales
using scaled and unscaled TICA followed by regular-space clustering
with 28 clusters each. }
\end{figure}

Fig. \ref{fig:hmm2}a shows the distribution of simulated points.
The slow transition occurs between the two interlaced 'T' motives
and requires a zigzag path. Thus, the reaction coordinate is highly
nonlinear in the given input coordinates and TICA cannot find a linear
combination that approximates the true reaction coordinate $\psi_{2}$
well. Consequently, both TICA coordinates are needed in order to resolve
the reaction coordinate. Unfortunately, TICA projects the stationary
process $\psi_{2}$ out by construction and the corresponding approximated
eigenvalue $\lambda_{2}$ becomes approximately zero. Since there
are only two eigenvalues in this example, the second coordinate is
lost. As a result, the kinetic map is poor and gives rise to a much
poorer MSM than the discretization of the unscaled TICA coordinates
(Fig. \ref{fig:hmm2}b).

Note that this example is extremely pathological and is only supposed
to show that there are rare combinations of poor choices in which
the current approach can break down. The kinetic map has one dimension
less than the original input space, as the first eigenfunction is
constant (in TICA this coordinate is projected out as a result of
removing the mean in the basis functions (\ref{eq:tica_basis_set})).
Since in our example TICA needed both input coordinates for a good
approximation of the reaction coordinate, a poor kinetic map was obtained.
When a third input coordinate is added, this problem disappears. Molecular
problems are usually high-dimensional and are thus not expected to
cause the observed problem. However, a general lesson from this example
is that kinetic map will only be a good approximation of the real
kinetic map $(\lambda_{i}\psi_{i})_{i=2,\,...,\,n}$, and thus lead
to near-optimal distance metric, if we have a good approximation of
the eigenvalues and eigenfunctions $(\hat{\lambda}_{i},\hat{\psi}_{i})$.
Although TICA and more general the method of linear variation \cite{NueskeEtAl_JCTC14_Variational,NoeNueske_MMS13_VariationalApproach}
only find linear combinations of input coordinates we can turn these
methods into excellent approximators of nonlinear eigenfunctions by
providing suitable input coordinates. In MD simulations, coordinates
such as interaction distances, contacts or torsion angles are expected
to play a role in the optimal reaction coordinates. Fortunately we
do not need to make a restrictive choice of coordinates, but can simply
add all promising coordinates to the input set and then run TICA or
the method of linear variation in order to find good combinations.

\paragraph{Bovine pancreatic trypsin inhibitor (BPTI)}

Let us turn to protein simulations. We analyzed a one-millisecond
simulation of BPTI produced by D. E. Shaw research using the Anton
supercomputer \cite{Shaw_Science10_Anton} (see there for simulation
setup). The trajectory was subsampled every 10 ns, providing 100,000
frames that are sufficient for our analysis. We then used the 174
coordinates of the 58 $C_{\alpha}$-atoms after aligning them to their
means as an input data set. We consider four metrics: (i) Euclidean
metric in the full TICA space, projection onto the first two (ii)
and the first six (iii) IC's, where gaps are found in the TICA timescales,
and (iv) the kinetic map of all scaled TICA-coordinates. $k$-means
clustering with 100 clusters was used in all cases.

\begin{figure}
\noindent \begin{centering}
a)\includegraphics[width=0.3\columnwidth]{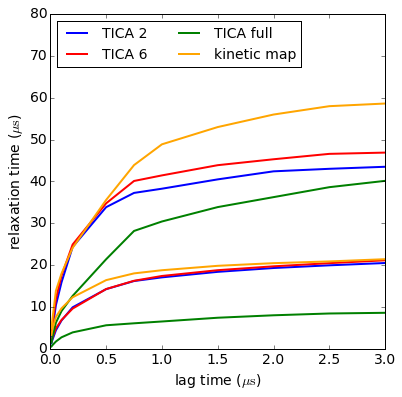} 
b)\includegraphics[width=0.3\columnwidth]{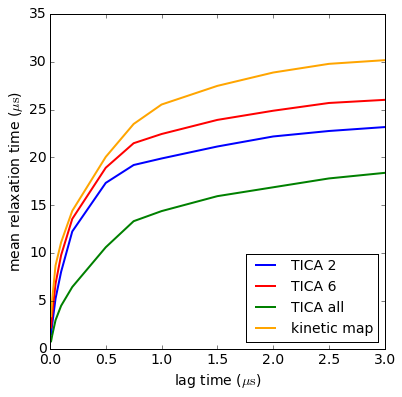}
c)\includegraphics[width=0.3\columnwidth]{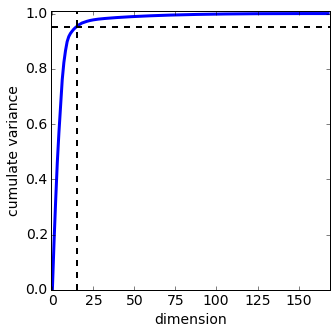}
\par\end{centering}

\noindent \begin{centering}
d)\includegraphics[width=0.8\columnwidth]{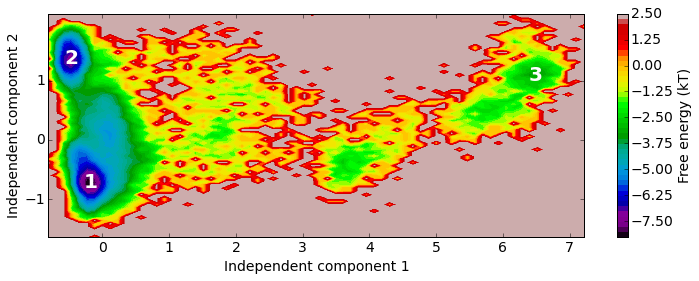}
\par\end{centering}

\noindent \begin{centering}
e.1) \includegraphics[width=0.2\columnwidth]{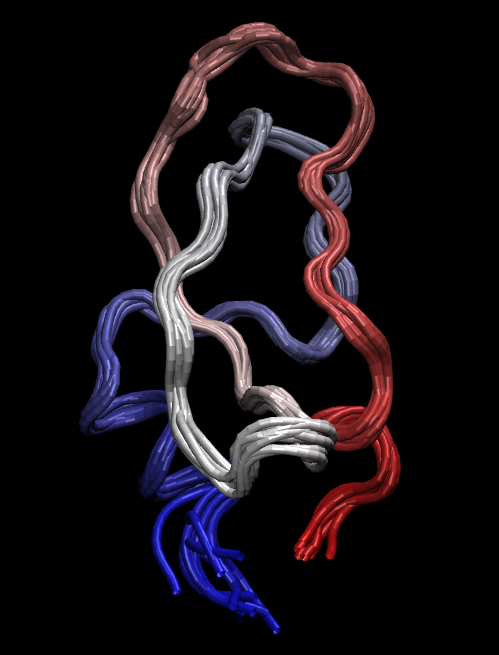}
e.2)\includegraphics[width=0.2\columnwidth]{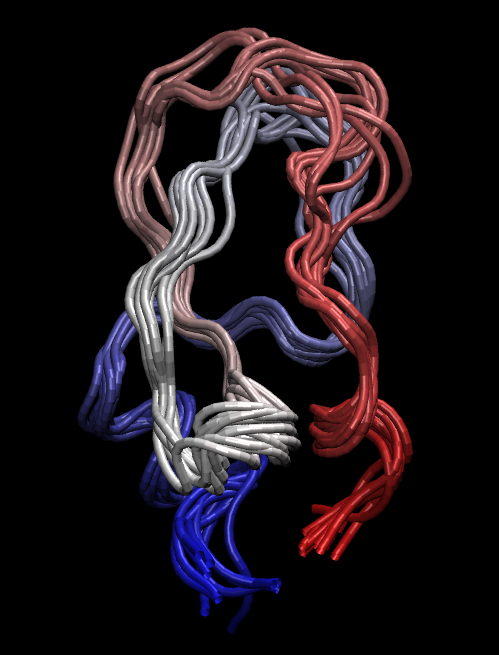}
e.3)\includegraphics[width=0.2\columnwidth]{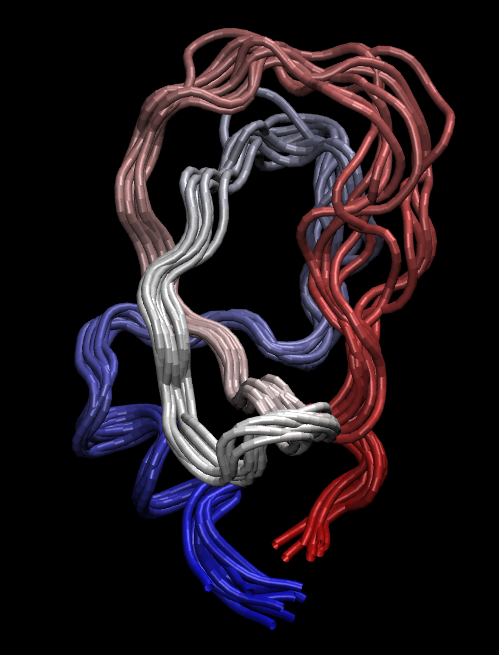} 
\par\end{centering}

\protect\caption{\label{fig:bpti}Comparison of Markov models of BPTI (1 millisecond
Anton trajectory \cite{Shaw_Science10_Anton}) using different TICA
projections and the kinetic map. TICA using the C$\alpha$ coordinates
of oriented BPTI configurations. MSMs built based on projections onto
2, 6 and all dimensions and the kinetic map were compared. (a) The
two slowest implied relaxation timescales. (b) Mean of the two slowest
relaxation timescales. (c) Cumulative variance of the diffusion distance.
95\% is reached by using 13 eigenvectors. (d) Kinetic map (first two
dimensions). (e.1-3) Structures of the metastable states indicated
in (d).}
\end{figure}

Fig. \ref{fig:bpti}a,b) show that the results using two and six IC's
are similar, but the MSMs built on all TICA coordinates are much worse.
The space is so high-dimensional that the clusters cannot efficiently
discretize the slow reaction coordinates. In contrast, the kinetic
map results are significantly better than all other setups, in particular
in terms of finding a converged slowest relaxation timescale. 

In order to get an idea of the effective dimensionality of the kinetic
map, Fig. \ref{fig:bpti}c shows the cumulative kinetic variance as
a function of the kinetic map dimension. 95\% of the variance are
obtained after only 13 dimensions, indicating that the data analysis
pipeline can work with relatively low-dimensional data after the TICA
step.

Fig. \ref{fig:bpti}d shows the first two dimensions of the BPTI kinetic
map with correctly scaled coordinates. Three metastable states are
apparent in this projection whose structures are depicted in Fig.
\ref{fig:bpti}e.1-3. The slowest conformational transition between
the pair (1,2) and state (3) (about 60 $\mu$s) involves an outward
motion of the loop around residue 10 (top right in the structure).
The second-slowest transition (about 20 $\mu$s) involves minor concerted
motions in the loop region and an exchange between an ordered set
of structures (1) and a less ordered set of structures (2). Qualitatively,
this analysis agrees with previous analyses of that system \cite{Shaw_Science10_Anton,NoeEtAl_PMMHMM_JCP13},
but the relaxation timescale found here are larger than previously
estimated. Following the variational principle this means a better
model was found here.

\paragraph{Trypsin-Benzamidine}

Finally, we analyzed protein-ligand association in Benzamidine and
Trypsin using 491 trajectories of length 100 ns each that have been
generated on GPUgrid \cite{BuchFabritiis_PNAS11_Binding} (see there
for simulation details). The trajectories were saved every 100 ps,
providing 491,000 frames for the analysis. As coordinates we chose
the distance between the $C_{7}$ atom of benzamidine with the $C_{\alpha}$
atoms of Trypsin, providing 223 distances. We consider four metrics:
(i) Euclidean metric in the full TICA space, projection onto the first
two (ii) and the first ten (iii) IC's, and (iv) the kinetic map of
all scaled TICA-coordinates. $k$-means clustering using 100 clusters
was used in all cases.

\begin{figure}
\noindent \begin{centering}
a)\includegraphics[width=0.3\columnwidth]{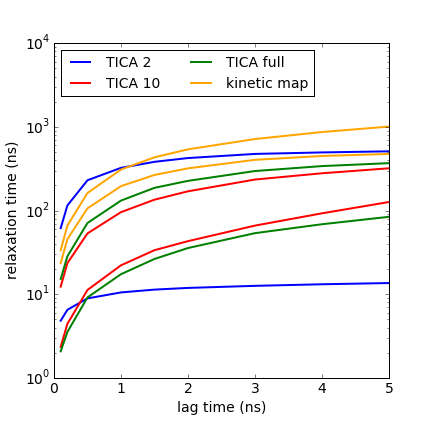}
b)\includegraphics[width=0.3\columnwidth]{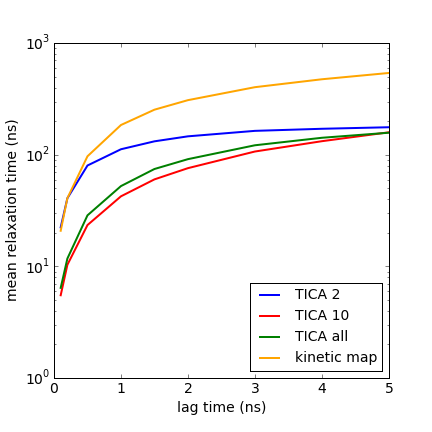}
c)\includegraphics[width=0.3\columnwidth]{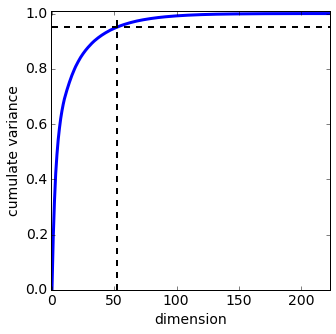}
\par\end{centering}

\noindent \begin{centering}
d) \includegraphics[width=0.5\columnwidth]{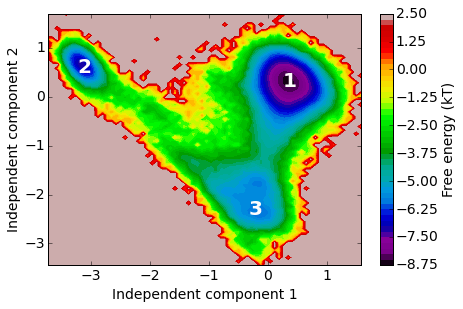}
e.1)\includegraphics[width=0.4\columnwidth]{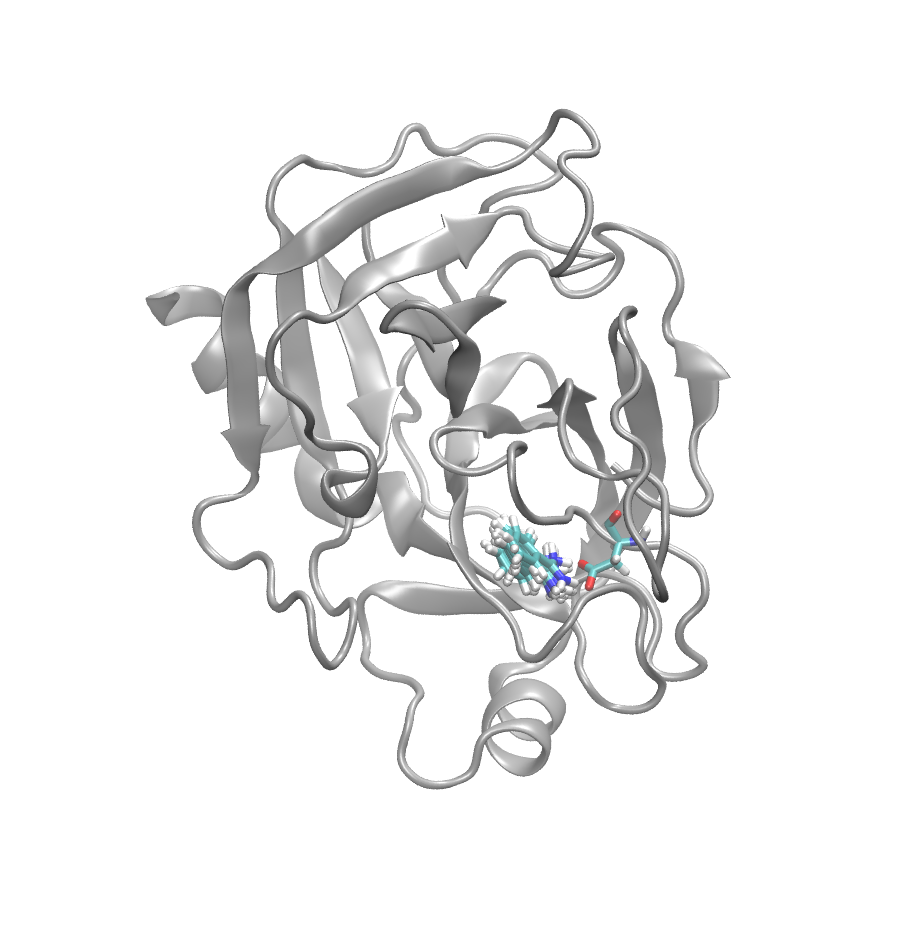}
\par\end{centering}

\noindent \begin{centering}
e.2)\includegraphics[width=0.4\columnwidth]{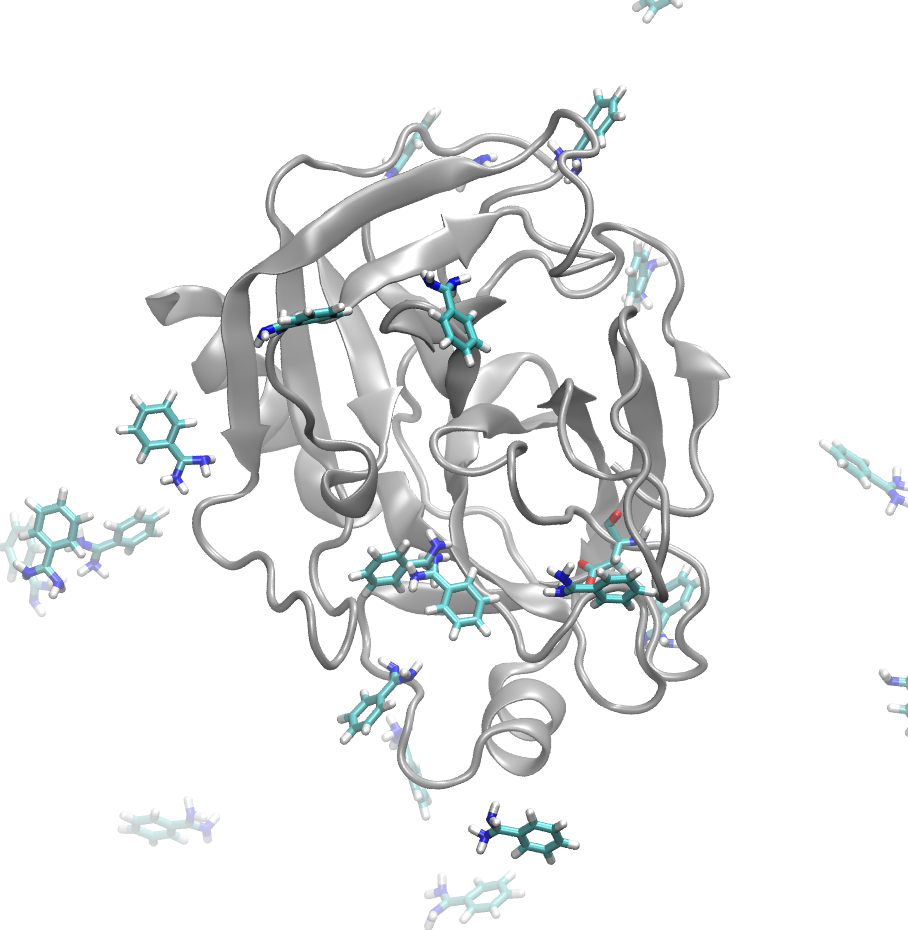}
e.3)\includegraphics[width=0.4\columnwidth]{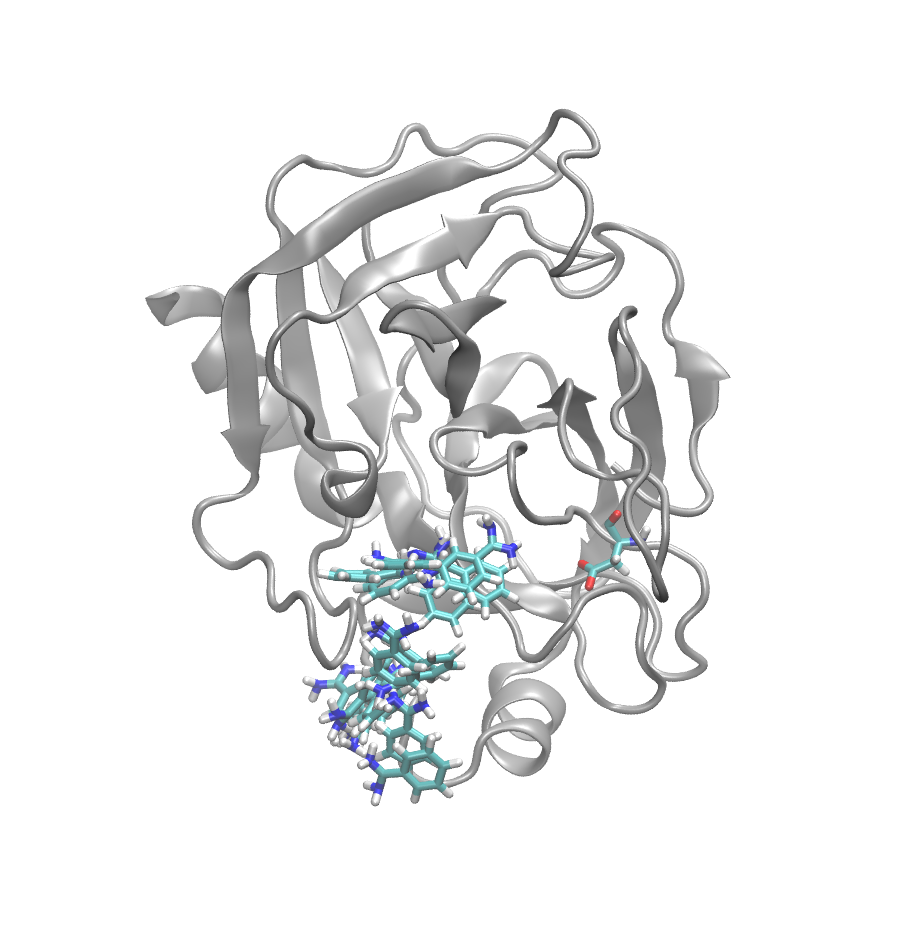}
\par\end{centering}

\protect\caption{\label{fig:trypsin}Comparison of Markov models of trypsin-Benzamidine
dynamics (491 GPUgrid trajectories of 100 ns each \cite{BuchFabritiis_PNAS11_Binding})
using different TICA projections and scaled TICA. TICA using the distances
from Benzamidin to all Trypsin-C$\alpha$ coordinates had 135 usable
eigenvalues. Projections onto 2, 5, 10 and all 135 dimensions (unscaled)
were compared to scaled TICA. (a) Implied relaxation timescales. (b)
Mean relaxation timescale. (c) Cumulative variance of the diffusion
distance. 95\% is reached by using 50 eigenvectors. (d) Kinetic map
(first two dimensions). (e) MD configurations sampled from the three
metastable states visible in d (1: bound, 2: dissociated, 3: pre-bound).}
\end{figure}

Fig. \ref{fig:trypsin}a,b) show that the MSMs using two dimensions
converge nicely to timescales of about 500 ns and 50 ns. For larger
number of ICs (10 and all) the results become worse. On one hand 100
clusters are no longer sufficient to discretize these higher-dimensional
spaces, and on the other hand it seems that a different second-slowest
process is found when looking at higher numbers of dimensions. The
kinetic map results show significantly larger timescales than any
of the other metrics. These timescales do not converge within the
range of lagtimes shown, but it is known from \cite{PlattnerNoe_NatComm15_TrypsinPlasticity}
that the Benzamidine coordinates relative to trypsin are actually
not sufficient to characterize the slowest processes in the system
which are comprised of trypsin conformational switches. 

Fig. \ref{fig:bpti}c shows the cumulative kinetic variance as a function
of the kinetic map dimension. 95\% of the variance are obtained after
52 out of 233 dimensions, indicating that the data can be reduced
by a factor of about 4-5 with little losses.

Fig. \ref{fig:bpti}d shows the first two dimensions of the trypsin-benzamidine
kinetic map with correctly scaled coordinates. Three metastable states
are apparent in this projection whose structures are depicted in Fig.
\ref{fig:bpti}e.1-3. According to the MSM on the input coordinates
used here, the slowest conformational transition is the binding unbinding
transition, although we know that slower transitions exist in the
protein conformation \cite{PlattnerNoe_NatComm15_TrypsinPlasticity}.
The second-slowest process involves exchange with a binding intermediate
where the ligand interacts with trypsin residues close to the binding
site. Qualitatively, this analysis agrees with previous analyses of
that system that used the trypsin coordinates \cite{BuchFabritiis_PNAS11_Binding},
but the kinetic map shows larger relaxation timescales. Following
the variational principle this means a better model was found here.

\section{Discussion}

The kinetic distance defined here is the optimal distance metric for
analyzing metastable molecular dynamics. In practice, the kinetic
distance can only be computed approximately and requires a method
that approximates the eigenvalues and eigenfunctions of the Markov
backward propagator underlying the MD simulation. Here we suggest
to use the time-lagged independent component analysis (TICA) as a
fast and convenient method to transform a large set of Cartesian or
internal coordinates of the molecule into such an approximation. Subsequently,
the kinetic distance can be computed. The TICA coordinates can be
transformed into a kinetic map by weighting them with their eigenvalues.
In this kinetic map, kinetic distances and Euclidean distances are
approximately equivalent, which makes it an excellent space for visualization
and further analyses such as clustering, Markov modeling, or diffusion
map.

We have shown that as long as pathological cases are avoided, the
kinetic distance provides a distance metric that builds better Markov
models, because the TICA components are weighted in such a way that
the clustering algorithm can optimally concentrate on the slow coordinates.
Instead of projecting onto an arbitrary number of TICA dimensions
as in previous work, or trying to select optimal values using machine
learning techniques, the present theoretical insights lead to a unique
and indisputable choice of using all coordinates in a weighted form.
To reduce computational effort, a controlled truncation of the TICA
space can be made by defining the percentage of the cumulative variation
in kinetic distance (e.g. 95\%).

The method can be improved by using methods that provide better approximations
of the true eigenvalues and eigenfunctions. Much can be done here
by choosing a more suitable set of input coordinates for the TICA
calculation, but a host of other related methods such as the Variational
Approach for conformation dynamics, Diffusion maps, and Markov transition
models are available that may further improve this estimate.

The approach described herein is implemented in the development version
of pyEMMA (available at github.com/markovmodel/pyemma), and will be
included in the next official release (see www.pyemma.org for download
instructions, documentation and examples). We have made kinetic maps
the default setting when computing a TICA transform.

\paragraph{Acknowledgements}

We thank Kyle Beauchamp (MSKCC New York), John Chodera (MSKCC New
York), Benedict Leimkuhler (University of Edinburgh) and Alessandro
Laio (SISSA Trieste) for inspiring discussions. Special thanks to
Ben for organizing a great workshop which was the starting point for
this paper. We are grateful to Gianni De Fabritiis for sharing the
trypsin-benzamidine data and D. E. Shaw research for sharing the BPTI
data. CC is supported by National Science Foundation (grants CHE-1152344
and CHE-1265929) and the Welch Foundation (grant C-1570). FN is supported
by European Commission ERC starting grant 307494-pcCell and Deutsche
Forschungsgemeinschaft (grant NO 825/3-1).

\clearpage

\end{document}